\documentstyle[prl,aps,preprint,psfig]{revtex}

\begin{document}
%@twocolumnfalse\endcsname
%%
%%
%\draft
\tighten
\title{Spectrum of Background X-rays from Moduli Dark Matter}
\author{T. Asaka}
\address{Institute for Cosmic Ray Research, University of Tokyo,
  Tanashi 188-8502, Japan}
\author{J. Hashiba}
\address{Department of Physics,  University of
  Tokyo, Tokyo 113-0033, Japan}
\author{M. Kawasaki}
\address{Institute for Cosmic Ray Research, University of Tokyo,
  Tanashi 188-8502, Japan}
\author{T. Yanagida}
\address{Department of Physics,  University of
  Tokyo, Tokyo 113-0033, Japan}
\date{\today}

\maketitle

\begin{abstract}
    We examine the $X$-ray spectrum from the decay of the dark-matter
    moduli with mass $\sim {\cal O}(100)$keV, in particular, paying
    attention to the line spectrum from the moduli trapped in the halo
    of our galaxy.  It is found that with the energy resolution of the
    current experiments ($\sim 10$\%) the line intensity is about
    twice stronger than that of the continuum spectrum from the moduli
    that spread in the whole universe. Therefore, in the future
    experiments with higher energy resolutions it may be possible to
    detect such line photons. We also investigate the $\gamma$-ray
    spectrum emitted from the decay of the multi-GeV moduli.  It is
    shown that the emitted photons may form MeV-bump in the
    $\gamma$-ray spectrum. We also find that if the modulus mass is of
    the order of 10 GeV, the emitted photons at the peak of the
    continuum spectrum loses their energy by the scattering and the
    shape of the spectrum is significantly changed, which makes the
    constraint weaker than that obtained in the previous works.

\end{abstract}

\clearpage

%%%%%%%%%%%%%%%%%%%%%%%%%%%%%%%%%%%%%%%%%%%%%
%%%%%%%%%%%%%%%%%%%%%%%%%%%%%%%%%%%%%%%%%%%%%
\section{Introduction}
%%%%%%%%%%%%%%%%%%%%%%%%%%%%%%%%%%%%%%%%%%%%%
%%%%%%%%%%%%%%%%%%%%%%%%%%%%%%%%%%%%%%%%%%%%%
%
Superstring theories\cite{Green}, which may be the most attractive
candidates to unify all known interactions including gravity,
have a number of flat directions, called moduli fields, in a large
class of classical ground states\cite{Green}. These moduli fields $\phi$
continuously connect infinitely degenerate supersymmetric vacua
and they are generally expected to acquire their masses $m_\phi$
of the order of the gravitino mass $m_{3/2}$ once supersymmetry
breaking effects are included\cite{Carlos-Casas-Quevedo-Roulet}.

These moduli fields cause different kinds of cosmological problems
\cite{Coughlan,K-Y} depending on values of their masses.
At present the thermal inflation proposed by Lyth and Stewart
\cite{Lyth-Stewart} seems to be the most plausible solution to the problems.
In recent articles\cite{H-K-Y,A-H-K-Y}, we have shown by postulating
the thermal inflation that only two regions of the moduli masses,
$m_\phi \lesssim $ 500 keV and $m_\phi \gtrsim {\cal O}$(100) GeV,
are cosmologically viable. In particular, the lighter mass region
is more interesting since the original Affleck-Dine baryogenesis
\cite{Affleck-Dine} does work here as shown first by
de Gauv\^{e}a, Moroi and Murayama\cite{G-M-M}. On the contrary,
for $m_\phi \gtrsim {\cal O}$(100) GeV we must invoke
a variant type of Affleck-Dine baryogenesis\cite{S-K-Y} which
has not been, however, well investigated yet.

If the moduli masses lie indeed in the region
$m_\phi \simeq 10^{-2}$ keV--200 keV there is an intriguing
possibility\cite{A-H-K-Y} that the moduli fields are
the dark matter in our universe. Since the thermal inflation
produces a tremendous amount of entropy at 
the late epoch of the universe's evolution to dilute the moduli density
substantially, there seems to be no candidate left for the dark matter
beside the moduli themselves
\footnote{
The axion with high values of decay constant
$F_a \simeq 10^{15}$--$10^{16}$ GeV could be another candidate
for the dark matter\cite{L-S-S-S}.
}.
This would encourage us to consider the hypothesis of moduli being
the dark matter in the universe.

In this paper we calculate spectrum of background X-rays emitted from
the moduli dark matter and find that the spectrum is constituted of
two distinct parts: one comes from the cosmic moduli filling
homogeneously the whole universe and the other from the moduli
condensed on the dark halo in our galaxy. The former has a relatively
broad spectrum due to the redshift effect and the latter has a peak
in the energy spectrum. We show that the peak in the X-ray spectrum
can be detectable in future experiments if the moduli masses $m_\phi$
are around 100 keV. We also briefly comment on $\gamma$-ray spectrum
emitted from more massive moduli of $m_\phi \simeq$ 1 -- 10GeV, since
this multi-GeV mass region is marginally allowed \cite{H-K-Y,A-H-K-Y}
if one assumes somewhat smaller values of the initial amplitudes of
moduli fields, $\phi_0 \simeq$ (0.01 -- 0.1)$M_G$, where $M_G$ is the
gravitational scale $M_G \simeq 2.4 \times 10^{18}$ GeV. We find that
the $\gamma$-rays emitted from such moduli make a large bump in
multi-MeV region and the shape of the spectrum depends heavily on the
masses of moduli.
%
%%%%%%%%%%%%%%%%%%%%%%%%%%%%%%%%%%%%%%%%%%%%%
%%%%%%%%%%%%%%%%%%%%%%%%%%%%%%%%%%%%%%%%%%%%%
\section{Cosmological moduli problem and the thermal inflation}
%%%%%%%%%%%%%%%%%%%%%%%%%%%%%%%%%%%%%%%%%%%%%
%%%%%%%%%%%%%%%%%%%%%%%%%%%%%%%%%%%%%%%%%%%%%
%
In this section we briefly review the previous works
\cite{H-K-Y,A-H-K-Y} and show that only two mass regions
such as $m_\phi \lesssim$ 500 keV and $m_\phi \gtrsim {\cal O}$(100) GeV
survive various cosmological constraints. We assume
$m_\phi \simeq m_{3/2}$ throughout this paper.

Let us start with the cosmological moduli problems.  As mentioned in
the introduction the moduli fields of masses in the range of keV-TeV
cause different kinds of cosmological problems
\cite{Coughlan,Carlos-Casas-Quevedo-Roulet,K-Y} depending on their
masses.  On one hand, the moduli with masses ${\cal O}$(100) GeV $
\lesssim m_\phi \lesssim {\cal O}$(1) TeV decay soon after
nucleosynthesis and in consequence spoil the success of big-bang
nucleosynthesis.  On the other hand, the moduli with masses $m_\phi
\lesssim$ 1 GeV are entangled with the problem that the abundance of
moduli themselves overcloses the present universe, or the radiation
expected from their decay may exceed the observed cosmic
X($\gamma$)-ray backgrounds.  In any case, one finds it difficult to
solve these problems as far as one is concerned with the standard
cosmology.

We give now a brief account of the thermal inflation which is thought
to be the only mechanism to overcome the above problems.  In their
original paper, Lyth and Stewart\cite{Lyth-Stewart} proposed a
solution for the case of moduli masses ${\cal O}$(100) GeV $ \lesssim
m_\phi \lesssim {\cal O}$(1) TeV, that corresponds to hidden sector
supersymmetry breaking models~\cite{Nilles}.  Thus, let us restrict
ourselves in the following discussion to the mass region $10^{-2}$ keV
$\lesssim m_\phi \lesssim$ 10 GeV, that is relevant to gauge-mediated
supersymmetry breaking
models~\cite{Giudice}.\footnote{%%
The detailed analysis has been carried out in \cite{H-K-Y}. For a more
complete version, see Ref.\cite{A-H-K-Y}.}

When we discuss the abundance of moduli energy density $\rho_\phi$,
it is convenient to introduce a ratio $\rho_\phi / s$, where
$s$ is the entropy density, since this quantity is invariant under
the universe's evolution as long as no entropy is produced.
Then the problems stated above are reexpressed as that
the present value of the ratio $\rho_\phi / s$
is predicted to be greater than the ratio $\rho_c / s$,
where $\rho_c$ is the critical density of the present universe,
by typical factors of $10^{11}$ -- $10^{16}$. If the thermal inflation
takes place, however, a significant increase in the entropy density
leads to an extreme reduction of the quantity $\rho_\phi / s$.

The existence of flaton fields is required to provide vacuum energy,
which is responsible for the thermal inflation to occur,
and to produce entropy by their decay into radiation.
The potential of the flaton $X$ is given by
\cite{Lyth-Stewart,H-K-Y}\footnote{%%
The imaginary part of the complex scalar field $X$ can be
interpreted as a massless NG boson originating from the U(1) symmetry
possessed by the potential (\ref{flaton-pot}). The NG bosons
produced by the flaton decay diminish drastically
the dilution effect of the thermal inflation.
We have given in \cite{A-H-K-Y} such a modification of the original
thermal inflation model that this unfavorable decay mode is suppressed.
It has been found, however, that the modification did not affect
the original dynamics in Ref.\cite{H-K-Y}}
\begin{equation}
    V = V_{0} - m_{0}^{2}|X|^{2} + \frac{1}{M_{*}^{2n}}
    |X|^{2n+4},
    \label{flaton-pot}
\end{equation}
where $-m_{0}^{2}$ is a negative mass squared induced by SUSY breaking
effect and $M_{*}$ a cut-off scale of this effective theory. the
vacuum energy density $V_{0}$ is determined so that the cosmological
constant vanishes at the true vacuum.

If the flaton $X$ couples to some fields which are in
thermal bath, the flaton potential (\ref{flaton-pot}) with
finite temperature effects taken into account reads
\begin{equation}
    V_{{\rm eff}} = V_{0} + (cT^2 - m_{0}^{2})|X|^{2}
    + \frac{1}{M_{*}^{2n}}|X|^{2n+4}.
    \label{flaton-eff-pot}
\end{equation}
Here, $T$ is the cosmic temperature and $c$ a constant of ${\cal O}$(1).
As be easily seen from the form of the effective potential
(\ref{flaton-eff-pot}), the flaton sits near the origin
at the temperature $T > T_c \simeq m_{0}$, and gives rise to the vacuum
energy density $V_{0}$. This vacuum energy becomes greater than
the radiation energy at the temperature
$T < T_* \simeq V_{0}^{1/4}$, since the radiation energy density
is given by $\rho_{{\rm rad}}=(\pi^{2}/30)g_{*}T^4$, where
$g_{*}$ is the effective number of degrees of freedom.
Therefore, for the temperature $T_c < T < T_*$ the flaton vacuum energy
density dominates the cosmic energy density, and the thermal inflation
takes place.

When the cosmic temperature becomes lower than the critical temperature,
i.e. $T < T_c$, the flaton begins rolling down towards the true minimum
of the potential (\ref{flaton-pot}), and oscillates around it.
The flaton coherent oscillation energy is eventually transferred
to the radiation energy through the flaton decay and reheats
the universe, increasing the entropy density by a factor of
\begin{equation}
         \Delta \simeq \frac{4V_{0}/3T_{R}}
        {(2\pi^2/45)g_{*}T_{c}^{3}}
        \simeq \frac{V_{0}}{70T_{R}T_{c}^{3}},
        \label{dilution}
\end{equation}
where $T_{R}$ is the reheating temperature.

We are now at the point to evaluate the moduli energy density,
with the notable effects of the thermal inflation considered.
We assume only one modulus $\phi$ to exist for simplicity.
The generalization to the case of many moduli is straightforward, however.

When the Hubble parameter $H$ becomes comparable to the modulus mass
$m_\phi$, the coherent oscillation of the modulus,
which we refer to as `big-bang modulus', starts
with the initial amplitude $\phi_0$ which is likely to be
of the order $M_G$. Then, the abundance of `big-bang modulus'
after the thermal inflation is calculated as
\begin{eqnarray}
    \left(\frac{\rho_{\phi}}{s}\right)_{{\rm BB}} &\simeq&
    \frac{m_{\phi}^{2}\phi_{0}^{2}/2}{8.6m_{\phi}^{3/2}M_{G}^{3/2}}
    \frac{1}{\Delta}, \nonumber\\ &\simeq& 4
    \left(\frac{T_{c}}{m_0}\right)^{3}
    \left(\frac{\phi_{0}}{M_{G}}\right)^{2}
    \frac{M_{G}^{1/2}m_{\phi}^{1/2}m_0^{3}T_{R}}{V_{0}}.
    \label{bb-moduli}
\end{eqnarray}
We should not forget `thermal inflation modulus' that is
a secondary oscillation, which begins after the thermal inflation,
stimulated by the shift
$\delta \phi \sim (V_{0}/m_{\phi}^2M_{G}^2)\phi_{0}$
from the true minimum of the modulus potential. The abundance of
this `thermal inflation modulus' is
\begin{eqnarray}
    \left(\frac{\rho_{\phi}}{s}\right)_{{\rm TI}} &\simeq&
    \frac{m_{\phi}^{2}(\delta\phi)^{2}/2}{(2\pi^2/45)g_{*}T_{c}^{3}}
    \frac{1}{\Delta}, \nonumber\\ &\simeq&
    \frac{3}{8} \left(\frac{\phi_{0}}{M_{G}}\right)^{2}
    \frac{V_{0}T_{R}}{m_{\phi}^{2}M_{G}^{2}}.
    \label{ti-moduli}
\end{eqnarray}
The total energy density of the modulus $\phi$ is then given by
\begin{equation}
    \frac{\rho_{\phi}}{s} \simeq
    \max \left[\left(\frac{\rho_{\phi}}{s}\right)_{{\rm BB}},
      \left(\frac{\rho_{\phi}}{s}\right)_{{\rm TI}}\right].
    \label{moduli-density}
\end{equation}

In Ref.\cite{H-K-Y}, we regarded $m_0$ and $M_*$ as free parameters
and obtained the theoretically predicted lower bound of
(\ref{moduli-density}), under the condition $T_{R} \gtrsim$ 10 MeV
that is required in order for the radiation created by the flaton decay
not to upset the nucleosynthesis. As a result, we have shown that
for all the region $10^{-2}$ keV $\lesssim m_\phi \lesssim$ 10 GeV
the lower bound of $\Omega_\phi h^2 \equiv \rho_\phi h^2/\rho_c$
($h$ is the present Hubble parameter $H_0$ in units of 100km/sec/Mpc)
could be taken below the critical density $\Omega h^2 \simeq 0.25$.

A constraint from the observed X($\gamma$)-ray backgrounds, however,
can be more stringent\cite{K-Y} than that from the critical density in
a certain modulus mass region. The modulus decays into two photons
dominantly.  Thus, we can derive another constraint on $\Omega_\phi
h^2$ by requiring that the maximum value of the predicted photon flux
should be less than the observed X($\gamma$)-ray backgrounds.  It has
been shown in \cite{H-K-Y} that this constraint excludes an
interesting mass region 500 keV $\lesssim m_\phi \lesssim$ 10 GeV.

Let us summarize the conclusions that were obtained in
\cite{H-K-Y,A-H-K-Y}.  First, we have found that only the theories
with modulus mass $10^{-2}$ keV $\lesssim m_\phi \lesssim$ 500 keV
could survive the cosmological constraints.%
\footnote{
If we take the cut off scale $M_\ast$ in Eq.(\ref{flaton-pot})
as $M_\ast \gtrsim M_G$,
which is a natural choice,
only the modulus with $m_\phi \sim$ 100 keV is
cosmologically allowed for $m_\phi \lesssim$ 500 keV
and becomes the dark matter of our universe
\cite{Asakaetal}.}
Second, we have pointed
out that the modulus with mass $m_\phi \simeq$ 1 -- 10 GeV also had a
chance to be allowed cosmologically if we could take $\phi_0 \simeq$
(0.01 -- 0.1)$M_G$. The final one, which has motivated us to work on this
paper, is that in the modulus mass region $10^{-2}$ keV $\lesssim
m_\phi \lesssim$ 200 keV the equality $\Omega_\phi h^2 \simeq {\cal
O}$(1) could be fulfilled \cite{A-H-K-Y} because in this region the
constraint from X($\gamma$)-ray backgrounds is weaker than that from
the critical density. This observation is none other than the reason
why we have stressed in the introduction that the moduli could be the
dark matter in our universe.

%%%%%%%%%%%%%%%%%%%%%%%%%%%%%%%%%%%%%%%%%%%%%
%%%%%%%%%%%%%%%%%%%%%%%%%%%%%%%%%%%%%%%%%%%%%
\section{X-ray Spectrum from Moduli Dark Matter}
%%%%%%%%%%%%%%%%%%%%%%%%%%%%%%%%%%%%%%%%%%%%%
%%%%%%%%%%%%%%%%%%%%%%%%%%%%%%%%%%%%%%%%%%%%%
%
As shown in previous section, the modulus field with mass $m_\phi
\simeq$ $10^{-2}$keV -- 200 keV is a candidate for the dark matter of our
universe.  This upper limit of the modulus mass originates from the
constraint of the observed cosmic photon backgrounds.  The modulus
field decays most likely to two photons through non-renormalizable
interaction suppressed by the
gravity scale.\footnote{%%
The modulus decay into two neutrinos is suppressed due
to the chirality flip.}
The lifetime of the modulus is estimated as\cite{K-Y,H-K-Y}
\begin{eqnarray}
    \label{tau-phi}
    \tau_\phi ~\simeq~ 
    \frac{ 64 \pi }{ b^2 } \frac{ M_G^2 }{ m_\phi^3 }
    ~ \simeq ~
    7.6 \times 10^{23} ~ \mbox{sec}~\frac{1}{b^2}
    \left( \frac{ 1 ~\mbox{MeV} }{ m_\phi } \right)^3,
\end{eqnarray}
where $b$ denotes a parameter of order one which depends on the models
of the superstring.  In the following we take $b$ = 1.  From
eq.(\ref{tau-phi}) the modulus with mass $m_\phi \lesssim$ 100 MeV has
a lifetime longer than the age of the present universe.  However, such
modulus is continuously decaying at the rate of $1/\tau_{\phi}$ and
produce photons which contribute to the diffuse photon backgrounds.
This excludes the region 500 keV $\lesssim m_\phi \lesssim$ 1 GeV.
Furthermore, when we assume the cosmic modulus field as the dark
matter of our universe ($\Omega_\phi \simeq {\cal O}$(1)), 
a region $10^{-2}$keV
$\lesssim m_\phi \lesssim$ 200 keV survives from the cosmological
constraints.

If the modulus field is indeed the dark matter, it would be the other
contribution to the photon backgrounds.  Some of the dark-matter
moduli should be trapped in the halo of our galaxy.  Since the Doppler
spread due to the velocities of the moduli is negligible in this case,
the narrow line spectrum is expected by the decay of the dark-matter
moduli in our halo.

In this section we consider the dark-matter moduli with mass $m_\phi
\sim$ 100 keV and investigate the $X$-ray spectrum of the produced
photons in the decay, since such $X$-ray may be observable in the
future experiments as we will describe below.

First we discuss the $X$-ray spectrum by the decay of the cosmic
moduli distributed uniformly over the whole universe.  Through the
modulus decay two monochromatic photons with energy $E_\gamma =
m_\phi/2$ are produced.  When we integrate such line spectrum from the
past to the present, the continuum spectrum below the energy $m_\phi /
2$ is observed today.  In Refs.\cite{K-Y,H-K-Y,A-H-K-Y} the photon
flux from the decay of the moduli in the whole universe is estimated
as
\begin{eqnarray}
    \label{F_U}
    F_U( E_\gamma ) ~\simeq~
    \frac{1}{4\pi}
    \frac{ 2 \Omega_\phi }{ \tau_\phi m_\phi}
    \frac{ \rho_c }{ H_0 }
    \left( \frac{ 2 E_\gamma}{m_\phi } \right)^{3/2}
    f(m_{\phi}/2E_{\gamma})
    \exp(-t(z=m_{\phi}/2E_{\gamma}-1)/\tau_{\phi}),
\end{eqnarray}
with
\begin{equation}
    f(x) = [\Omega_0 + (1-\Omega_0 - \Omega_{\lambda})/x
    + \Omega_{\lambda}/x^3]^{-1/2},
\end{equation}
where $\Omega_{\lambda}$ is the density parameter of the cosmological
constant, $z$ the redshift and $t(z)$ the cosmic time at $z$. 
This flux takes its maximal value at the photon energy
\begin{eqnarray}
    E_{\rm max} ~\simeq~ \left\{\begin{array}{ll}
          \frac{ m_\phi }{ 2 } 
          &\mbox{for}~~ \tau_{\phi} > t(z=0)\\
          \frac{ m_{\phi} }{ 2 } 
          \left( \frac{ 3 \tau_{\phi} H_{0}\sqrt{\Omega_0}}{ 2 } 
          \right)^{2/3}
          ~~~~~~~
          &\mbox{for}~~ \tau_{\phi} < t(z=0)
          \end{array}\right..
\end{eqnarray}
In particular, when the modulus field is the dark matter, $\tau_{\phi}
\gg  t(z=0)$ and $F_U(E_{\rm max})$ is given by
\begin{equation}
    F_U(E_{\rm max}) =     \frac{1}{4\pi}
    \frac{ 2 \Omega_\phi }{ \tau_\phi m_\phi}
    \frac{ \rho_c }{ H_0 }.
\end{equation}
It should be notice that this equation is independent of $\Omega_0$
and $\Omega_{\lambda}$.

In Fig.1 we show the spectrum of the photon flux (\ref{F_U}) for
various moduli masses.  Then the $X$-ray intensity from the whole
universe is given by
\begin{eqnarray}
    \label{continuos}
    I_U(E_\gamma)~\simeq~
    \frac{1}{E_\gamma} F_U(E_\gamma).
\end{eqnarray}
As shown in Fig.1, the flux from the moduli decay is comparable to the
observed $X$-ray backgrounds if the mass of the modulus is $\sim 100$keV.

Next we consider the $X$-ray spectrum coming from the dark-matter
moduli trapped in the our galactic halo. Since the cosmological
redshift is negligible in this case, the photons produced by the decay
have a monochromatic energy $m_{\phi}/2$. Here we estimate the
intensity of this line spectrum.  The mass density of the halo of our
galaxy at the distance $r$ from the center of the galaxy is expressed
as\cite{Halo-Density}
\begin{eqnarray}
    \label{rho_halo}
    \rho_H (r) ~\simeq~
    \frac{ \rho_0 }{ 1 + \frac{ r^2 }{ r_c^2 } },
\end{eqnarray}
where $r_c \simeq 2$ kpc and the halo density in the solar
neighborhood ($r \simeq R_0 \simeq$ 8.5 kpc) is $\rho_H(R_0) \simeq$
0.38 GeV/cm$^3$.  Then the density of the modulus component in the halo
is given by $\rho_H(r) \times ( \Omega_\phi / \Omega_0 )$.  Using the
halo density (\ref{rho_halo}) the line flux is estimated as
\begin{eqnarray}
    F_H ~\simeq~
    \frac{ 1 }{ 4 \pi }
    \frac{ 2 }{ \tau_\phi m_\phi }
    \int dx
    \frac{ \Omega_\phi }{ \Omega_0 }
    \frac{ \rho_0 r_c^2 }
    { ( x - R_0 \cos b \cos l )^2 
     + R_0^2 ( 1 - \cos^2 b \cos^2 l ) 
     + r_c^2 },
\end{eqnarray}
where $x$ is the distance to the modulus particle from the sun and $l$
($b$) is the galactic longitude (latitude).  After the $x$
integration, we obtain the following expression
\begin{eqnarray}
    F_H ~\simeq~
    \frac{ 1 }{ 4 \pi }
    \frac{ 2 }{ \tau_\phi m_\phi }
    \frac{ \Omega_\phi }{ \Omega_0 }
    \frac{ ( R_0^2 + r_c^2 ) \rho_H(R_0) }{ R_{eff} }
    \left[ 
        \frac{ \pi }{2}
        + \tan^{-1} 
          \left( \frac{ R_0 \cos b \cos l }{ R_{eff} }
          \right)
    \right],  
\end{eqnarray}\
with $R_{eff}^2 = R_0^2 ( 1 - \cos^2 b \cos^2 l ) + r_c^2$.
Then the diffuse line intensity from the galactic halo is
given by
\begin{eqnarray}
    I_H ~\simeq~ \frac{F_H}{ \Delta E},
\end{eqnarray}
where $\Delta E$ denotes the energy resolution at $E_\gamma \simeq
m_\phi/2$ of the experiment.  Here it should be noted that the line
flux depends on the direction of the incoming photon, i.e. $b$- and
$l$-dependence, and that the intensity of the line $X$-ray spectrum
from the galactic halo becomes more significant in the experiments with
higher energy resolution, which contrasts to the continuous spectrum
produced by the decay of the moduli in the whole universe.

Then we compare the line intensity from the moduli in our galactic
halo to the maximum value of the $X$-ray intensity from the moduli
that spread over the whole universe.  For this end it is convenient to
introduce the ratio $R_I$ defined as
\begin{eqnarray}
    \label{ratio-intensity}
    R_I(b,l) &=& I_H/I_U(E_{\rm max}).
\end{eqnarray}
For the dark-matter moduli this ratio $R_I$ is almost independent on
the modulus mass $m_\phi$ since we can neglect the exponential factor
in eq.(\ref{F_U}).  If we see the direction of the north or south
galactic poles, this ratio becomes
\begin{eqnarray}
    R_I(b=\pm \pi/2, l)
    ~\simeq~
    0.16 \frac{ 1 }{ \Omega_0 } 
    \left( \frac{ E_{\rm max} }{ \Delta E } \right).
\end{eqnarray}

In Fig.2 we show the contour of the ratio $R_I$ in the $b$-$l$
plane for the case $m_\phi \simeq$ 200 keV (i.e. $E_\gamma$ = 100 keV)
and $\Delta E/E$ = 10 \%.\footnote{%%
The $X$-ray backgrounds at energy $E_\gamma \simeq 100$ keV were
measured by $HEAO$-I experiment\cite{HEAO-I} whose energy resolution
$\Delta E/E$ is about 10 \%.  }
We find that the line intensity is about two times stronger than the
peak of the continuum spectrum in the wide region of the sky.  In
Fig.3 we also show the intensity of the line photons in the direction
$b=\pi/2$ together with the continuous spectrum (\ref{continuos}) for
the energy resolution $\Delta E/E=10$\% and 5\% and for the modulus
mass 100keV and 200keV. It is seen that the photon intensity from the
moduli in the whole universe is below or marginal to the observed one
for $m_{\phi} = 100$keV or 200keV, but the line intensity from the
halo moduli is above the observed one.  Thus, future experiments with
energy resolution $\Delta E/E \lesssim 10$\% at energy around
$E_\gamma = m_\phi/2$ can detect the line intensity from the
dark-matter moduli in our halo for $m_\phi \gtrsim 100$keV. 
Furthermore, by observing the $b$- or
$l$-dependence of the $X$-ray intensity of the peak, we may confirm
the origin of the peak, i.e. it originates from the line spectrum
produced by the decay of the dark-matter moduli in the halo of our
galaxy.
%
%%%%%%%%%%%%%%%%%%%%%%%%%%%%%%%%%%%%%%%%%%%%%
%%%%%%%%%%%%%%%%%%%%%%%%%%%%%%%%%%%%%%%%%%%%%
\section{$\gamma$-ray Spectrum from Cosmic Multi-GeV Moduli}
%%%%%%%%%%%%%%%%%%%%%%%%%%%%%%%%%%%%%%%%%%%%%
%%%%%%%%%%%%%%%%%%%%%%%%%%%%%%%%%%%%%%%%%%%%%
%
The cosmic modulus field with multi-GeV mass decays into two photons
until the present. Since the produced photons are redshifted by the
cosmic expansion, they form the continuum spectrum which takes it
maximal value at MeV region.  Thus the decay of the multi-GeV modulus
field may be observed as a MeV-bump in the spectrum of the background
$\gamma$-rays if the produced photons reach us directly. However, such
high energy photons may be scattered off the background photons and
its spectrum may be deformed.  For the case of the multi-GeV modulus,
we can neglect the double photon pair creation process: $\gamma +
\gamma_{BG} \rightarrow e^{+} e^{-}$ because the energy of the
produced photons is below the effective threshold $E_{*} \simeq
m_e^2/(22T)$($m_e$: electron mass, $T$: background
temperature)~\cite{Kawasaki}. Thus we take into account only the
photon photon scattering process: $\gamma + \gamma_{BG} \rightarrow
\gamma + \gamma$ by which the emitted photons from the moduli lose
their energy. Since the total cross section of the photon photon
scattering is proportional to $E_\gamma^3$, ($E_\gamma$ denotes an
energy of the emitted photon.), this process becomes significant only
for modulus with mass larger than $O(1)$GeV.  In this section we
estimate the photon spectrum emitted from the multi-GeV moduli
including the effect of the scattering with the background's photons.

In order to obtain the photon spectrum we solve the following
Boltzmann equation for the distribution function
$f_\gamma$~\cite{Svensson}:
\begin{eqnarray}
    \frac{ \partial f_\gamma(E_\gamma) }{ \partial t }
    &=& 
    \frac{1112}{10125 \pi} \frac{\alpha^4}{m_e^8}
    \int_{E_\gamma}^{\infty} d \epsilon_\gamma ~
    f_\gamma(\epsilon_\gamma) \epsilon_\gamma^2
    \left[ 1 - \frac{E_\gamma}{\epsilon_\gamma}
           + \left(\frac{E_\gamma}{\epsilon_\gamma}\right)^2
    \right]^2
    \int_0^{\infty} d \overline{\epsilon} ~
    \overline{\epsilon}^3 ~\overline{f}(\overline{\epsilon})
\nonumber \\
    &-& 
    \frac{1946}{50625 \pi} \frac{\alpha^4}{m_e^8}
    E_\gamma^3 f_\gamma(E_\gamma)
    \int_0^\infty d \overline{\epsilon} ~
    \overline{\epsilon}^3 ~\overline{f}(\overline{\epsilon})
\nonumber \\
    &-&
    2 H f_\gamma(E_\gamma)
\nonumber \\
    &+&
    \frac{1}{4 \pi} \frac{ 2 \rho_{\phi} }{ \tau_\phi m_\phi }
    \mbox{e}^{ - \frac{t}{\tau_\phi} }
    \delta \left( E_\gamma - \frac{m_\phi}{2} \right),
    \label{bz-eq}
\end{eqnarray}
where $\overline{f}$ denotes the distribution function
of the background photon at temperature $T$:
\begin{eqnarray}
    \overline{f}(\epsilon) = \frac{\epsilon^2}{\pi^2} \times
    \frac{ 1 }{ \exp(\epsilon/T) - 1 }.
\end{eqnarray}
We solve the Boltzmann equation (\ref{bz-eq}) numerically including
the evolution of the universe.  The continuum $\gamma$-ray spectrum of
the modulus decay is obtained using the present distribution function
as $I_{U}(E_{\gamma})$ = $\left. f_{\gamma}(E_{\gamma}) \right|_{t =
t_{0}}$.
 
We show the spectra for $\Omega_{\phi}h^2 = 1$ and $m_{\phi} = 1, 10$
and $20$GeV in Fig.4.  The effect of the photon photon
scattering off the background photons is negligible for $m_{\phi} \sim
1$GeV.  On the other hand, for the modulus with mass $m_{\phi} \gtrsim
10$ GeV, we find that photons at the peak of the spectrum
significantly lose their energy by the scattering and the peak of the
spectrum moves to a lower energy region.  Therefore, comparing with
the observed background photon spectrum, it is found that the
constraint becomes slightly weaker for the modulus with mass $m_{\phi}
= {\cal O}(10)$ GeV than that obtained in Ref.\cite{A-H-K-Y}.

%%%%%%%%%%%%%%%%%%%%%%%%%%%%%%%%%%%%%%%%%%%%%
%%%%%%%%%%%%%%%%%%%%%%%%%%%%%%%%%%%%%%%%%%%%%
\section{Conclusion}
%%%%%%%%%%%%%%%%%%%%%%%%%%%%%%%%%%%%%%%%%%%%%
%%%%%%%%%%%%%%%%%%%%%%%%%%%%%%%%%%%%%%%%%%%%%
%
In this paper we have examined the photon spectra from the decay of
the cosmic modulus field.  First we have considered the modulus mass
region $m_{\phi} \simeq$ $10^{-2}$keV--200 keV.  This region is
interesting because the modulus field can be the dark
matter in our universe.  We have calculated the $X$-ray continuum
spectrum from the decay of the dark-matter moduli that spread
homogeneously in the whole universe and the line spectrum from the
dark-matter moduli trapped in the halo of our galaxy.  It is found
that with the energy resolution of the current experiments ($\sim
10$\%) the line intensity is about twice stronger than that of the
continuum spectrum in the wide region of the sky.  If the modulus mass
is around 100 keV, both intensities are comparable with the present
observed photon backgrounds.  Therefore, in the future experiments
with higher energy resolutions it may be possible to detect the line
photons produced by the decay of dark-matter moduli in our halo.
Moreover, by measuring the dependence of the line intensity on the
galactic longitude and latitude, we will be able to confirm the
origin, i.e. it comes from the halo of our galaxy rather than from the
whole universe.

We have also investigated the $\gamma$-ray spectrum emitted from the
decay of the multi-GeV modulus field.  In this modulus mass region,
the emitted photons are redshifted and have a peak in the MeV region
of the spectrum. Thus we may observed those photons as a MeV-bump in
the $\gamma$-ray backgrounds. 

The produced high energy photon may be scattered off the background
photons and lose their energy.  It is found that the effect of the
scattering is negligible for modulus with mass less than ${\cal
O}(1)$GeV.  However, if the modulus mass is of the order of 10 GeV,
the emitted photons at the peak of the continuum spectrum loses their
energy by the scattering and the shape of the spectrum is
significantly changed.  This makes the constraint from the present
observed $\gamma$-ray backgrounds weaker than the result in
Ref.\cite{A-H-K-Y}.

\acknowledgments

We would like to thank T. Kamae for useful comments and encouragement.

%%%%%%%%%%%%%%%%%%%%%%%%%%%%%%%%%%%%%%%%%%%%%%%%%%%%%%%%%%%%%%%%%%%%
%%%%%%%%%%%%%%%%%%%%%%%%%%%%%%%%%%%%%%%%%%%%%%%%%%%%%%%%%%%%%%%%%%%%
%%%%% ** Reference ** %%%%%%%%%%%%%%%%%%%%%%%%%%%%%%%%%%%%%%%%%%%%%%
%%%%%%%%%%%%%%%%%%%%%%%%%%%%%%%%%%%%%%%%%%%%%%%%%%%%%%%%%%%%%%%%%%%%

%%%%%%%%%%%%%%%%%%%%%%%%%%%%%%%%%%%%%%%%%%%%%%%%%%%%%%%%%%%%%%%%%%%%
%%%%%%%%%%%%%%%%%%%%%%%%%%%%%%%%%%%%%%%%%%%%%%%%%%%%%%%%%%%%%%%%%%%%
%%%%% ** Figure Caption ** %%%%%%%%%%%%%%%%%%%%%%%%%%%%%%%%%%%%%%%%%
%%%%%%%%%%%%%%%%%%%%%%%%%%%%%%%%%%%%%%%%%%%%%%%%%%%%%%%%%%%%%%%%%%%%
%%% Fig. 1
\begin{figure}
    \centerline{\psfig{figure=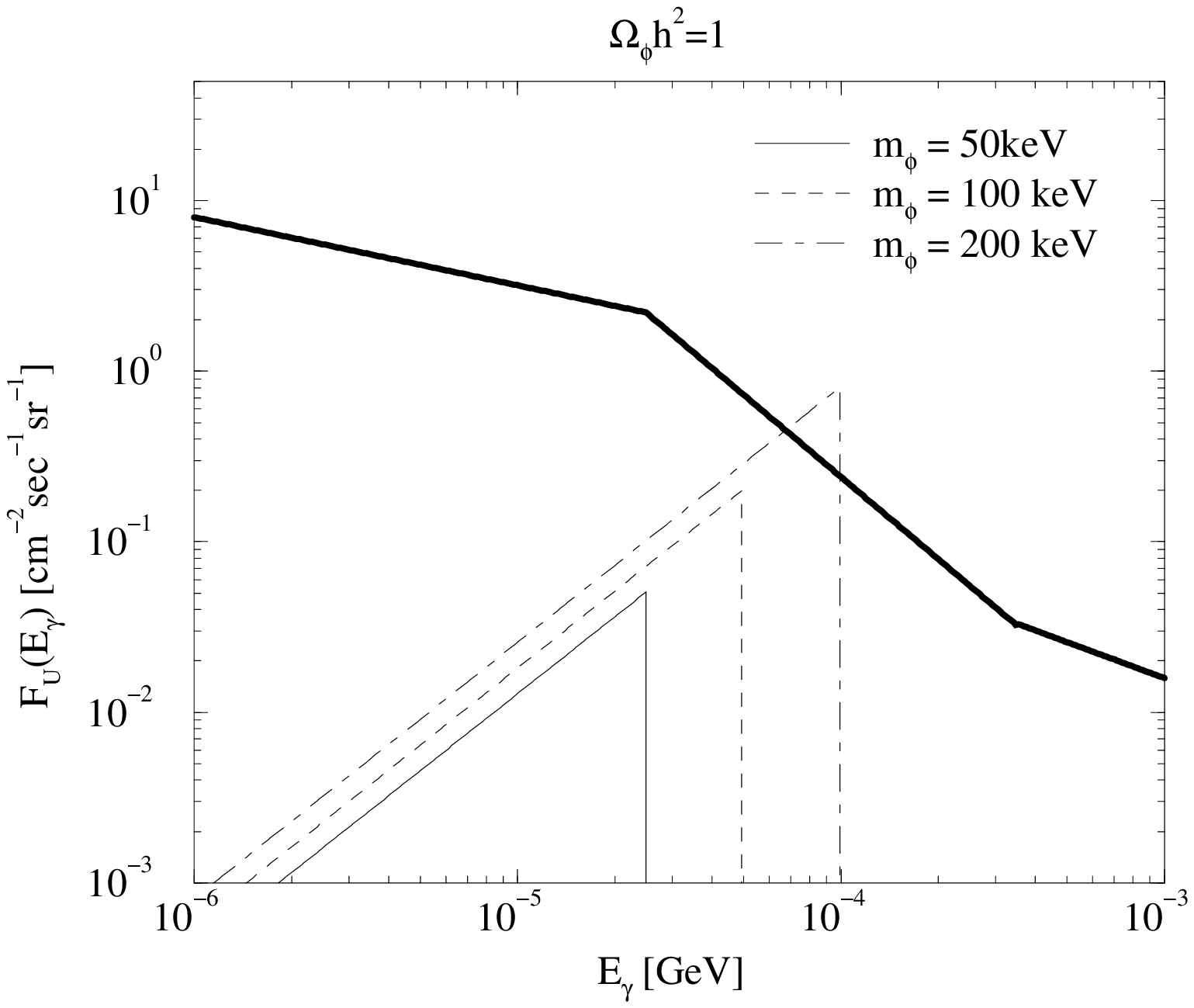,width=20cm}}
    \vspace{0.5cm}
\caption{%%
The spectra of the photon flux from the decay
of the cosmic moduli filling all the universe for various
moduli masses. We take $\Omega_\phi h^2$ = 1.
We also show the observed spectrum of the background photons
by the thick solid line.
}
\end{figure}
%%% Fig. 2
\begin{figure}
    \centerline{\psfig{figure=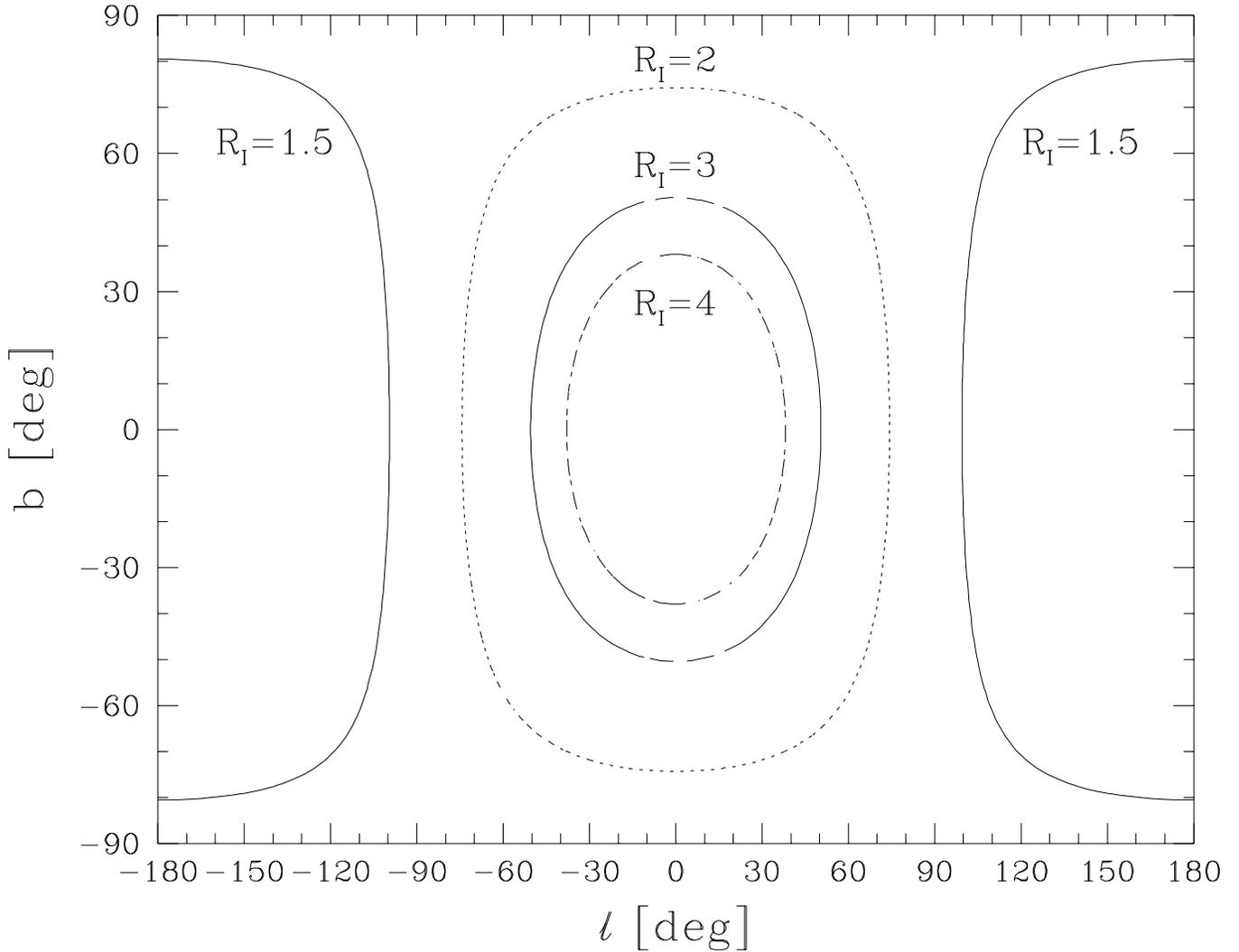,width=20cm}}
    \vspace{0.5cm}
\caption{%%
The contours of the ratio of the line intensity from
our galactic halo moduli to the maximum value of the continuum 
spectrum from the whole universe moduli.  We use the galactic
coordinates $(b,l)$. We assume the energy
resolution $\Delta E/E = 10$\% and $m_{\phi}=200$keV.
}
\end{figure}
%%% Fig. 3
\begin{figure}
    \centerline{\psfig{figure=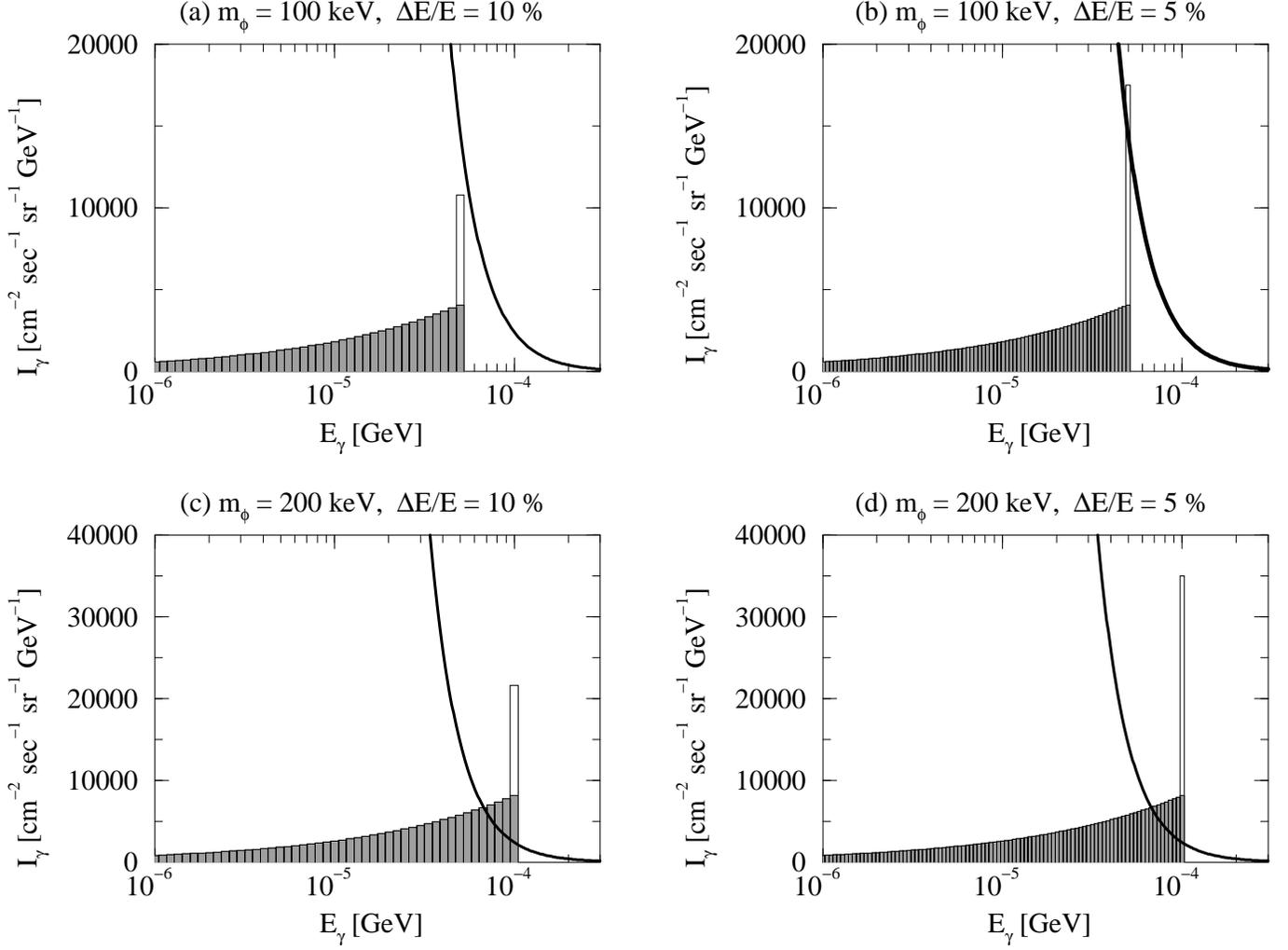,width=20cm,angle=270}}
    \vspace{0.5cm}
\caption{%%
The predicted intensity of the photons from the moduli in our halo
(white region) and from the moduli in the whole universe (dark region)
for the modulus mass $m_{\phi}=100$keV and 200keV. We assume that the
energy resolution $\Delta E/E = 10$\% and $\Delta E/E = 5$\%. For the
line spectrum of the our halo, we take the galactic latitude
$b=\pi/2$. We also show the observed spectrum of the background
photons by the thick solid line. }
\end{figure}
\clearpage
%%% Fig. 4
\begin{figure}
    \centerline{\psfig{figure=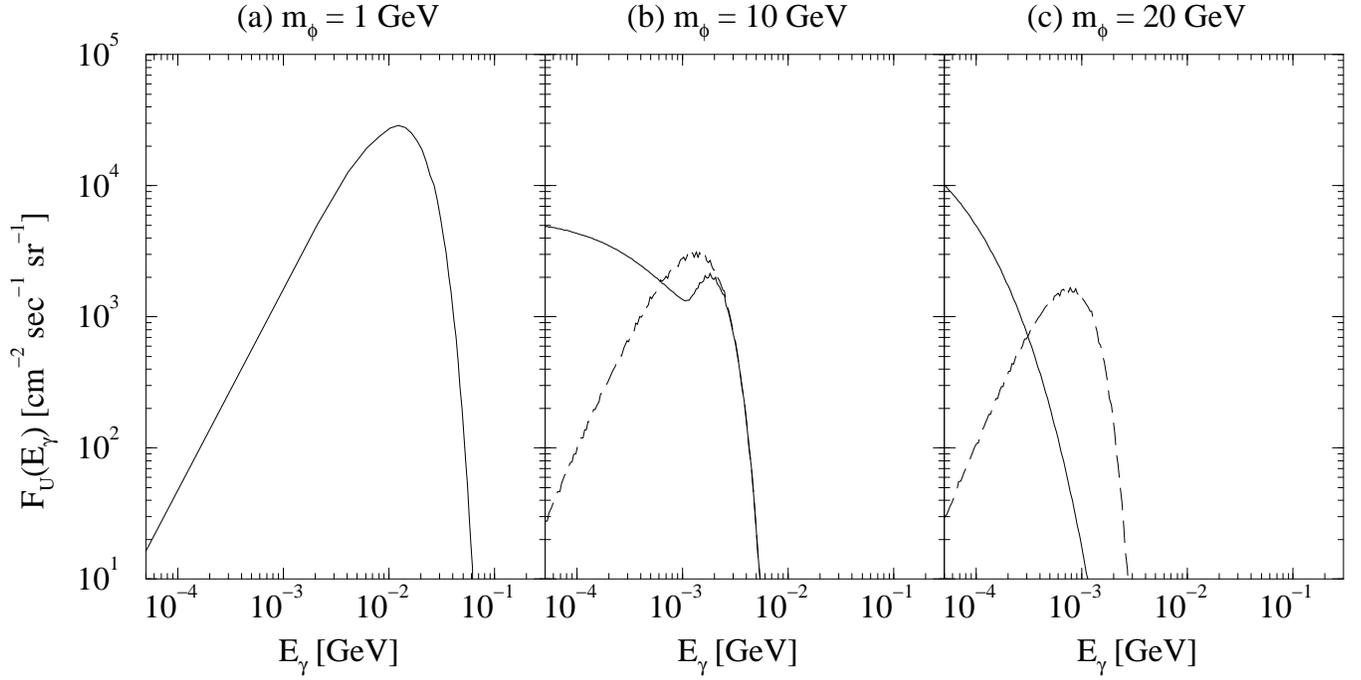,width=20cm,angle=270}}
    \vspace{0.5cm}
\caption{%%
The photon spectra from the decay of the modulus with mass $m_\phi$ =
1 GeV(a), $m_\phi$ = 10 GeV(b) and $m_\phi$ = 20 GeV(c).  The solid or
dot-dashed line represents the case with or without the effect of the
photon-photon scattering.  We take $\Omega_\phi h^2$ = 1.  }
\end{figure}
\end{document}